\documentclass[traditabstract]{aa}

\usepackage{txfonts}
\usepackage{graphicx}
\usepackage{natbib}
\usepackage{xspace}
\bibpunct{(}{)}{;}{a}{}{,}

\def\kMpc{\,h~\rm{Mpc}^{-1}}

\def\Mpc{\,h^{-1}~\rm{Mpc}}

\newcommand{\dif}{{\mathrm d}}

\usepackage{color,ulem}

\begin{document}

\title{
Accurate fitting functions for peculiar velocity spectra in standard and massive-neutrino cosmologies}



\author{
 J.~Bel\inst{\ref{cpt},\ref{brera}}
\and A.~Pezzotta\inst{\ref{barc1},\ref{barc2}, \ref{brera}}
\and C.~Carbone\inst{\ref{unimi},\ref{brera},\ref{infn-mi}}
\and E.~Sefusatti\inst{\ref{oats},\ref{brera},\ref{infn-ts}}
\and L.~Guzzo\inst{\ref{unimi},\ref{brera},\ref{infn-mi}}
}

\institute{
Aix Marseille Univ, Universit\'e de Toulon, CNRS, CPT, Marseille, France \label{cpt}
\and INAF - Osservatorio Astronomico di Brera, 
via E. Bianchi 46, 23807 Merate, Italy \label{brera}
\and Institute of Space Sciences (ICE, CSIC), Campus UAB, Carrer de Magrans s/n,
08193 Barcelona, Spain \label{barc1}
\and Institut d’Estudis Espacials de Catalunya (IEEC), 08034 Barcelona, Spain \label{barc2}
\and Universit\`{a} degli Studi di Milano, via G. Celoria 16, 20133 Milano, Italy \label{unimi}
\and INAF - Osservatorio Astronomico di Trieste, Via Tiepolo 11, 34143, Trieste, Italy\label{oats}
\and INFN - Sezione di Trieste, Via Valerio 2, 34127 Trieste, Italy\label{infn-ts}
\and INFN - Sezione di Milano, via G. Celoria 16, 20133 Milano, Italy \label{infn-mi}
}


\offprints{\mbox{J.~Bel}, \email{jbel@cpt.univ-mrs.fr}}

\abstract{We estimate the velocity field in a large set of $N$-body simulations including massive neutrino particles, and measure the auto-power spectrum of the velocity divergence field as well as the cross-power spectrum between the cold dark matter density and the velocity divergence. We perform these measurements at four different redshifts and within four different cosmological scenarios, covering a wide range in neutrino masses. We find that the nonlinear correction to the velocity power spectra largely depends on the degree of nonlinear evolution with no specific dependence on the value of neutrino mass. We provide a fitting formula based on the value of the r.m.s. of the matter fluctuations in spheres of $8h^{-1}$Mpc, describing the nonlinear corrections with 3\% accuracy on scales below $k=0.7\; h$ Mpc$^{-1}$.}


\maketitle

\section{Introduction}

The analysis of the large-scale structure of the universe provides crucial information on the evolution of the background matter density and its perturbations \citep[see, {\it e.g.,}][]{bernardeau02, Bassett2009, WeinbergEtal2013}. In particular, analyses of cosmological probes sensitive to different cosmic epochs 
could lead to an explanation of the mysterious late-time acceleration of cosmic expansion \citep[e.g.,][]{EuclidTheory}. At the same time, large-scale structure is sensitive to the details of the standard model of particle physics, providing upper bounds on the neutrino mass scale \citep[see, {\it  e.g.,}][]{L&P2006}. 

Measured redshifts, which are used to estimate galaxy distances, 
are affected by galaxy peculiar velocities generated by the growth of cosmological matter perturbations. Their
line-of-sight component
combines with the cosmological expansion, systematically modifying the derived galaxy distances and generating what are known as {\it redshift space distortions} (RSD). This effect turns the amplitude and isotropy of redshift-space clustering statistics into a sensitive probe of the 
linear growth rate of structure $f=d\ln{D}/d\ln{a}$ 
\citep{ka87}. 

Combining measurements of the expansion history $H(z)$ and the growth rate of structure $f$ can evidence deviations from the standard theory of gravity, that is, General Relativity. This \citep{GuzzoEtal2008} has led to renewed interest in RSD over the past decade \citep[see e.g.,][for the most recent analyses and a summary of previous results]{Sanchez2016,PezzottaEtal2017}.

Extracting the linear growth rate from RSD measurements of biased, nonlinear tracers as galaxies however requires an accurate modeling of redshift-space galaxy clustering.  Extensive work over  the past decade has addressed 
this in the context of cosmological perturbation theory, both in the Eulerian and Lagrangian formulations, providing linear and nonlinear predictions for the redshift-space galaxy power spectrum \citep[see, {\it e.g.,}][]{ka87, Scoccimarro2004, Matsubara2008a, Matsubara2008b, TNSH2009, TN&S2010, R&W2011, S&M2011, SeljakMcDonald2011, Valageas2011, Zhang2013, Zheng2013, TaruyaNishimichiBernardeau2013, S&Z2014, UK&H2015, Okumura20015, Perko2016, B&K2016, Bianchi2016, VlahCastorinaWhite2016, HandEtal2017A, HashimotoRaseraTaruya2017, FonsecaDeLaBella2017A}. 
\citet{Upadhye2016} in particular investigate RSD, taking into account massive neutrinos and dark energy.

In the formulation of \citet{Scoccimarro2004}, and several that followed, the large-scale effect of RSD is described in terms of two main ingredients:  the auto-power spectrum of the velocity divergence field $P_{\theta\theta}$, and the cross-spectrum between the velocity divergence and the matter density contrast $P_{\delta\theta}$. Although, perturbation theory appears to be a powerful tool to predict these quantities in the quasi-linear regime, it presents severe limitations when extended to smaller scales. In fact, RSD are characterised by a peculiar coupling of large- and small-scale clustering that represents a severe challenge to perturbative methods. Several assumptions usually made in the perturbative treatment, such as the irrotational nature of the velocity field, particularly relevant in the RSD modeling, are clearly not valid on small scales. 

An insight into the nonlinear evolution of the velocity field is offered
by numerical simulations, the standard tool for these kind of investigations. However, while 
extensive literature is dedicated to the description of the small-scale power spectrum and the characterisation of virialized structures as dark matter halos, the estimation of the velocity field in $N$-body simulations presents peculiar challenges.
This is due to the challenging estimation of the velocity field in low-density regions, resulting in a relatively limited number of studies on this specific topic
\citep[see, {\it e.g.,}][]{B&W1996, BernardeauEtal1997, P&S2009, Jennings2011, KodaEtal2014, ZhengZhangJing2015, ZhangZhengJing2015, Yu2015, HA&A2015, JenningsBaughHatt2015}. 

In particular, 
\citet{Jennings2012}, updating the previous results of \citet{Jennings2011}, provides a fitting formula for the nonlinear auto- and cross-power spectra of the velocity divergence, $P_{\theta\theta}$ and $P_{\delta\theta}$, calibrated using cosmological $N$-body simulations, in terms of the nonlinear matter power spectrum $P_{\delta\delta}$. Through a four-parameter fit, they reach a 2\% accuracy on $P_{\theta\theta}$ for $z=0$ and on scales $k< 0.65\kMpc$; the fit is less accurate for $P_{\delta\theta}$, and at higher redshifts. A similar but simpler fit for $P_{\delta\theta}$, again as a function of  $P_{\delta\delta}$ is proposed by \citet{Zheng2013}. It is expected to be 2\% accurate for all scales (and redshift) where the adimensional matter power spectrum  $\Delta_{\delta\delta}(k)\equiv 4\pi k^3P_{\delta\delta}(k)<1$. An alternative fit of the relation of $P_{\theta\theta}$ and $P_{\theta\delta}$ with 
$P_{\delta\delta}$ is represented by the simple exponential damping proposed instead by  \cite{HA&A2015}, which reaches a 5\% accuracy for $k<1\kMpc$.  

Any accurate modeling of galaxy clustering 
aiming at percent precision on the recovered  cosmological parameters 
should also account for
a nonzero neutrino mass. There are two main reasons for this. 
On one side, cosmological observations currently provide the best upper limits on the sum of neutrino masses \citep{Planck2016,PalanqueDelabrouilleEtal2015} and proper modeling is therefore required to extract the most unbiased estimates. On the other hand, at these levels of precision, neglecting the presence of this sub-dominant component would potentially add a comparable systematic error on any constraint derived from galaxy-clustering data 
\citep[see {\it e.g.,} ][]{BaldiEtal2014}. These are the motivations behind the significant effort spent  over the past few years to model 
the effect of neutrino masses on large-scale structure observables,
in particular using numerical simulations \citep[see, {\it e.g.,}][]{AliHaimoudBird2013, VillaescusaEtal2014, Castorina2015, Carbone2016, InmanEtal2015, TianNu2016, ZennaroEtal2017, VillaescusaNavarroEtal2017A, LiuEtal2017A}.

In this paper we use the Dark Energy and Massive Neutrinos Universe (DEMNUni) set of $N$-body simulations \citep{
Carbone2016} 
to model the velocity power spectra ($P_{\delta\theta}$ and $P_{\theta\theta}$) both in a standard ${\Lambda}$CDM scenario and including a massive neutrino component. The DEMNUni runs represent some of best simulations in massive neutrino cosmologies, both in terms of volume and mass 
resolution \citep{Castorina2015}. 
We 
propose 
an accurate fitting formula involving a minimal number of free parameters, which we calibrate against simulations, showing that the main dependence on cosmology can be encapsulated in a dependence on the current {\it r.m.s.} clustering amplitude $\sigma_{8}$ of the {\it cold} matter, that is, CDM and baryons. We focus on this component, as opposed to the total matter including neutrinos, as this appears to provide the simplest description of halo abundance and bias \citep{CastorinaEtal2014}. 

 
In Sect.  \ref{section1} we present our set of $N$-body simulations while in Sect.  \ref{section2} we describe the method used to estimate the cold dark matter velocity field. In Sect. \ref{section3} we discuss previous results obtained in past literature and compare them to our results, and in Sect. \ref{sec:summary} we summarise our findings in.

\section{Simulations}
\label{section1}

The DEMNUni   simulations  are a set of $N$-body cold dark matter simulations produced with the aim of testing multiple cosmological probes in the presence of massive neutrinos and dark-energy scenarios beyond the standard $\Lambda$CDM. They represent a reliable tool for exploring the impact of neutrinos on a wide range of dynamical scales, and have been extended to scenarios including a dynamical dark-energy background, with different equations of state parameters ($w_0$, $w_a$), in order to study
their degeneracy with the total neutrino mass at the nonlinear level. The technical implementation and detailed features of the simulations are presented in the description paper by \citet{Carbone2016} and  
the analysis of cold dark matter clustering by \citet{Castorina2015}. 

\begin{table}
{\renewcommand{\arraystretch}{1.5}
\begin{tabular}{cccccc}
$\sum m_\nu\,[\mathrm{eV}]$ & $\Omega_{cdm}$ & $\sigma_{8,mm}$ & $\sigma_{8,cc}$ & $\frac{m_p^c}{10^{10}}\,\bigg[\frac{M_\odot}{h}\bigg]$ & $\frac{m_p^\nu}{10^9}\,\bigg[\frac{M_\odot}{h}\bigg]$\\
\hline
\hline
$0.00$ & $0.2700$ & $0.846$ & $0.846$ & $8.27$ & $-$\\
$0.17$ & $0.2659$ & $0.803$ & $0.813$ & $8.16$ & $1.05$\\
$0.30$ & $0.2628$ & $0.770$ & $0.786$ & $8.08$ & $1.85$\\
$0.53$ & $0.2573$ & $0.717$ & $0.740$ & $7.94$ & $2.28$\\
\end{tabular}}
\vspace{0.5cm}
\caption{Parameters of the four $\nu\Lambda$CDM simulations that change among the different realisations. 
}
\label{table1}
\end{table}

In this work we exploit the first set of simulations, DEMNUni-I, describing several flat cosmological models characterised by various values of the total neutrino mass, $\sum m_\nu=0,0.17,0.3,0.53\;\mathrm{eV}$, while keeping the total matter density parameter fixed at $\Omega_m=0.32$. This implies that the cold dark matter relative density $\Omega_{cdm}$ changes across the four simulations in order to keep the sum $\Omega_{cdm}+\Omega_\nu=\Omega_m$ constant. The neutrino density is related to the total neutrino mass as $\Omega_\nu=\sum m_\nu/93.14/h^2\;\mathrm{eV}$ \citep{L&P2006}. Further properties shared by the four cosmologies are the density parameter associated to the cosmological constant $\Omega_\Lambda=0.68$ and to the baryon density $\Omega_b=0.05$, the Hubble constant $H_0=67\,km\, s^{-1}$Mpc$^{-1}$, the primordial spectral index $n_s=0.96$ and, most importantly, the scalar amplitude of the matter power spectrum $A_s=2.1265\times10^9$. As a consequence, while in the large-scale limit the power spectra of cold dark matter tend to the same value in all cosmological models, the value of the r.m.s. of cold dark matter perturbations on spheres of radius $8\Mpc$ depends on $\sum m_\nu$. The latter is denoted as $\sigma_{8,c}$, to distinguish it from the r.m.s of  total matter perturbations, $\sigma_{8,m}$.

The simulations were run on the FERMI supercomputer at CINECA\footnote{http://www.cineca.it/} ($5\times10^6$ CPU hours) using the tree particle hydrodynamical code GADGET-$3$ modified to include massive neutrino particles by \citet{Vieletal2010}. The latter regulates the assembly of $N_{cdm}=2048^3$ cold dark matter particles and $N_\nu=2048^3$ neutrino particles (when present) within a cubic periodic universe of comoving size $L=2000\;h^{-1}$Mpc. The mass resolution of cold dark matter and neutrinos varies slightly over the four simulations (values are listed in Table \ref{table1}), but in all cases it is large enough as to properly describe clustering in the nonlinear regime within the systematic error induced by neglecting baryonic effects. Initial conditions were set at redshift $z_{in}=99$ using the \citet{Zeldovich1970} approximation and were evolved to $z=0$, with a softening length $\varepsilon=20\;h^{-1}$kpc. During the runs, 62 snapshots were saved for each simulation, with equal logarithmic interval in the scale factor.

\section{Measurements of the density and velocity spectra}
\label{section2}

\begin{figure}
\centering
\includegraphics[width=.5\textwidth]{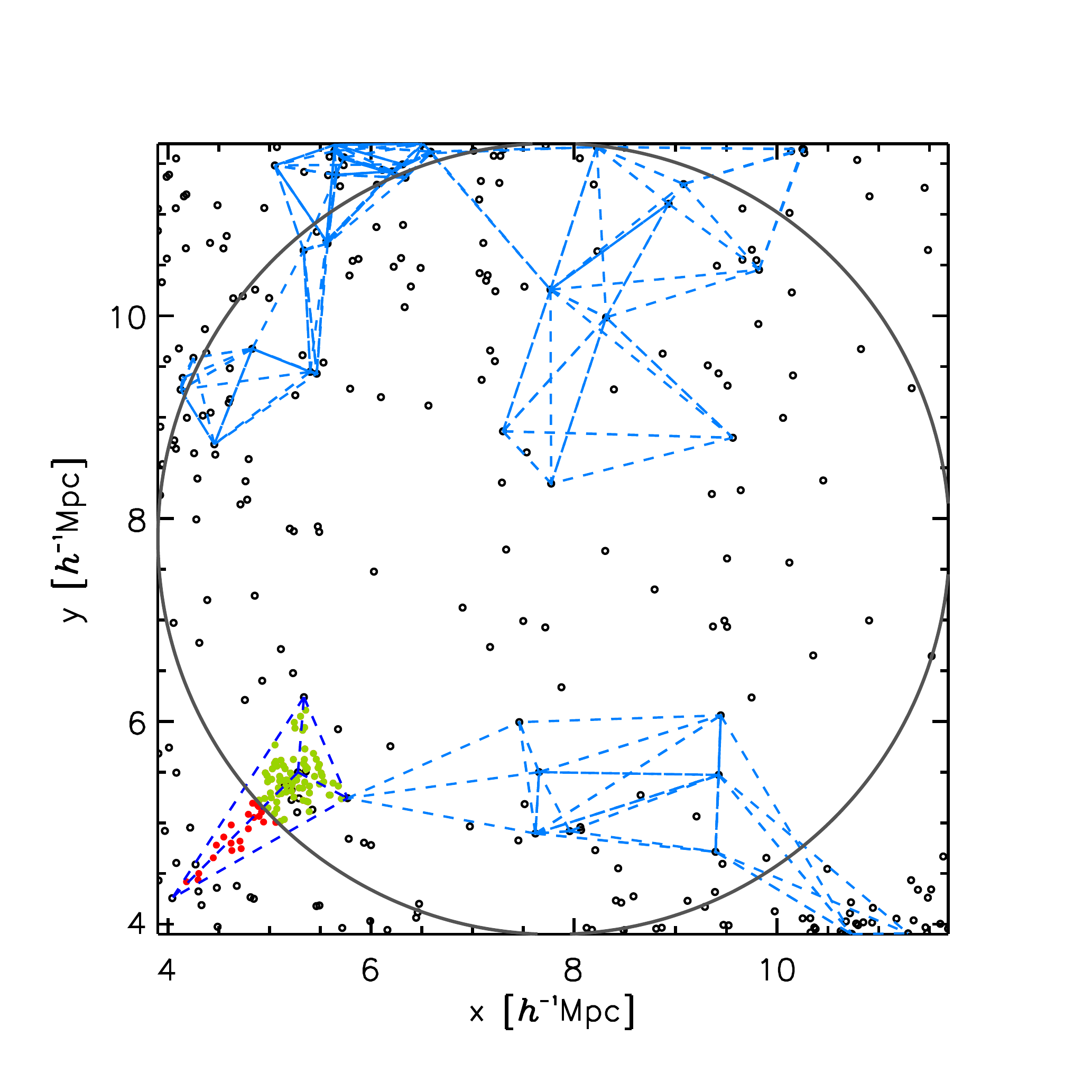}
\caption{Tetrahedra in a spherical cell: the empty circles represent the $3$D distribution of CDM particles in a cubic region of size $8h^{-1}$Mpc projected in $2$D. They represent the input set of points on which we perform the Delaunay tessellation. The gray circle represents the spherical cell. For clarity, we only show, in dashed light blue lines, the tetrahedra which have their four vertices included in a slice of $1.5h^{-1}$Mpc (in the $z$-direction) around the center of the cell. As an example we highlight one tetrahedron overlapping the boundary of the spherical cell (dark blue in the bottom left-hand corner). The red and green filled circles represent the randomly distributed points over the tetrahedron, the green ones represent those points lying inside the spherical cell and that are then used to assign a velocity to the corresponding region of the tetrahedron. This is repeated for all tetrahedrons of the tessellation that overlap the boundary of the spherical cell.
}
\label{fig:balls}
\end{figure}

The estimation of the velocity field in cosmological $N$-body simulations 
has been
investigated by \citet{B&W1996, BW&H1997, W&B1998, P&S2009}, or more recently with the Delaunay Tessellation Field Estimator by \citet{RW2007} and the Kriging method by \citet{Yu2015}. 
Here we adopt a method close to the one proposed by \citet{P&S2009}, which implements the Delaunay tessellation 
to reconstruct the velocity field from test particles of cold dark matter. We apply the count-in-cell technique to average both the velocity and density fields within spheres of radius $R=3.9\,h^{-1}$Mpc. Given the large number of particles in the simulation, we perform the Delaunay tessellation only locally around each cell, rather than applying it to the whole set of cold dark matter particles. We estimate the velocity field on a grid, dividing the simulation into $1024^3$ cubes of size $1.95h^{-1}$Mpc, which are used to index the particle positions.  

In order to estimate the velocity within a given spherical cell, we start by considering the particles belonging to the eight sub-cubes forming a cubical cell that contains the spherical cell itself. We then count how many particles are contained within the sphere; if it is greater than $200$ 
we choose to run the Delaunay tessellation 
over the cell particles inside the sphere, otherwise we run the Delaunay tesselation inside the cubical cell. In any case we
ensure that the total volume covered by the Voronoi tetrahedra inside the spherical cell includes at least $93$\% of the cell volume, otherwise we automatically extend the radius in which we are keeping the particles to perform the Delaunay tesselation by a factor $3/2$. We note that in practice the spherical cell size in unchanged. Only the effective volume of the Delaunay tesselation is varied in order to make sure that the volume fraction of tetrahedra within the spherical cell is representative at the $7$\% level of the volume of the cell.

The difficulty arising from the estimation of the velocity field convolved with our spherical (top-hat) window function is related to the treatment of the tetrahedra which are lying on the boundary of each spherical cell. 
Our method consists in weighting the average velocity of a tetrahedron by its volume if it lies entirely within the spherical cell. Instead, for tetrahedra that extend outside of 
the cell boundary, we generate a set of $100$ uniformly distributed random points inside the considered tetrahedron and assign them a velocity, linearly interpolated from those at each node of the tetrahedron. The same random points are also used to estimate the volume fraction of the tetrahedron lying inside the spherical cell. 
In this way, a velocity can be assigned to that specific sector of the spherical cell by averaging the velocities of the included random points.  
The result is then weighted by the corresponding volume fraction. This process is described in more detail in the following paragraphs and illustrated in Figure \ref{fig:balls}.

Each Voronoi tetrahedron can be described by a set of four points $\vec x_1$,  $\vec x_2$,  $\vec x_3$ and  $\vec x_4$ where $\vec x_i = (x_i, y_i, z_i)$; they are therefore defined by the transformation matrix \citep[see][]{P&S2009}
\begin{equation}
\Psi =
\left (
\begin{array}{ccc}
\Delta x_2 & \Delta x_3 & \Delta x_4 \\
\Delta y_2 & \Delta y_3 & \Delta y_4 \\
\Delta z_2 & \Delta z_3 & \Delta z_4 \\
\end{array}
\right ),
\end{equation}
where $\Delta x_i = x_i - x_1$. As a result, the volume of the $j$-th tetrahedron can be computed as $w_j = |\Psi|/6$. The matrix $\Psi$ can also be seen as the matrix transforming the basis of the vectors composed of $\Delta x_2$, $\Delta x_3$ and $\Delta x_4$ into the cartesian basis. If the position $\vec s$ is taken as the tetrahedron basis, then the corresponding coordinates in the cartesian basis are obtained as $\vec x = \vec x_1 + \Psi \vec s$. The same can be applied in order to interpolate linearly the velocity of a point located in $\vec s$, 
\begin{equation}
\vec v (\vec x) = \vec v_1 + \Phi \vec s ,
\label{velo}
\end{equation}
where 
\begin{equation}
\Phi =
\left (
\begin{array}{ccc}
\Delta v_{x,2} & \Delta v_{x,3} & \Delta v_{x,4} \\
\Delta v_{y,2} & \Delta v_{y,3} & \Delta v_{y,4} \\
\Delta v_{z,2} & \Delta v_{z,3} & \Delta v_{z,4} \\
\end{array}
\right ).
\end{equation}
For tetrahedra entirely contained inside the sphere one can show that the volume average of the interpolated velocity field inside the tetrahedron is the arithmetic mean of the four velocities taken at each vertex; $\vec V_j = \frac{1}{4}\sum_{i=1}^{4} \vec v_i $. As mentioned above, for tetrahedra crossing the  cell boundary, we randomly populate their volume with a uniform distribution of $N=100$ points (see Fig. \ref{fig:balls}) and we evaluate their corresponding velocities using equation \ref{velo}. The volume-averaged velocity assigned to the tetrahedron can be computed as
\begin{equation}
\vec V_j = \frac{1}{N_{\rm{in}}}\sum_{i=1}^{N_{\rm{in}}} \vec v(\vec x_i),
\label{vj}
\end{equation}
where $N_{\rm{in}}$ is the number of random points belonging to the spherical cell and the fraction of its volume inside the spherical cell can be estimated as $w_j = \frac{N_{\rm{in}}}{N}|\Psi|/6$. Finally, the volume-averaged velocity assigned to the spherical cell is obtained with the sum
%
\begin{equation}
\vec V = \frac{\sum_{j=1}^{N_t} w_j \vec V_j }{\sum_{j=1}^{N_t} w_j},
\end{equation}
where $N_t$ is the total number of tetrahedra with at least one vertex belonging to the spherical cell. We note that we checked explicitly that in the most clustered catalogue, that is, the one corresponding to the $\Lambda$CDM cosmology at $z=0$, the number of objects in a spherical cell is always greater than $8$, thus avoiding 
empty cells.

Once the velocity and density grids of $512^3$ regularly spaced sampling points have been built, we can Fourier transform them by means of a Fast Fourier Transform algorithm. Since a simple count-in-cell density interpolation can be severely affected by aliasing when transforming to Fourier space, we employ an interlacing  technique to reduce this spurious contribution \citep{H&E1988, Sefusatti2016}. Regarding the shot noise correction, we neglect it because the mean number of particles in each cell is $\bar N = 268$ which corresponds to a $3$\% contribution to the variance at $z=1.5$ and for the $M_{\nu}=0.53$eV (snapshot having the lowest variance). We then compute the divergence of the velocity field $\theta_{\bf k}$ by simply combining the three velocity grids as $\theta_{\bf k} = i( k_xv_x + k_yv_y +k_zv_z )$. Then the density power spectrum $P_{\delta\delta}$, the cross power spectrum $P_{\delta\theta}$ , and the divergence of the velocity power spectrum $P_{\theta\theta}$ are estimated by averaging over spherical shells in $k$-space. 
We note that we also average the modes, and assign the value of the angular average of the spectra to the $k$-space position of the corresponding mode average.

\section{Results}
\label{section3}

Our goal is to provide accurate prescriptions to estimate the $P_{\theta\theta}$ and $P_{\delta\theta}$ auto- and cross-spectra 
in the regime where the perturbative approach fails in describing the velocity field and its power spectra \citep{CS&B2012}. 
\begin{figure*}
\centering
\includegraphics[width=.9\textwidth]{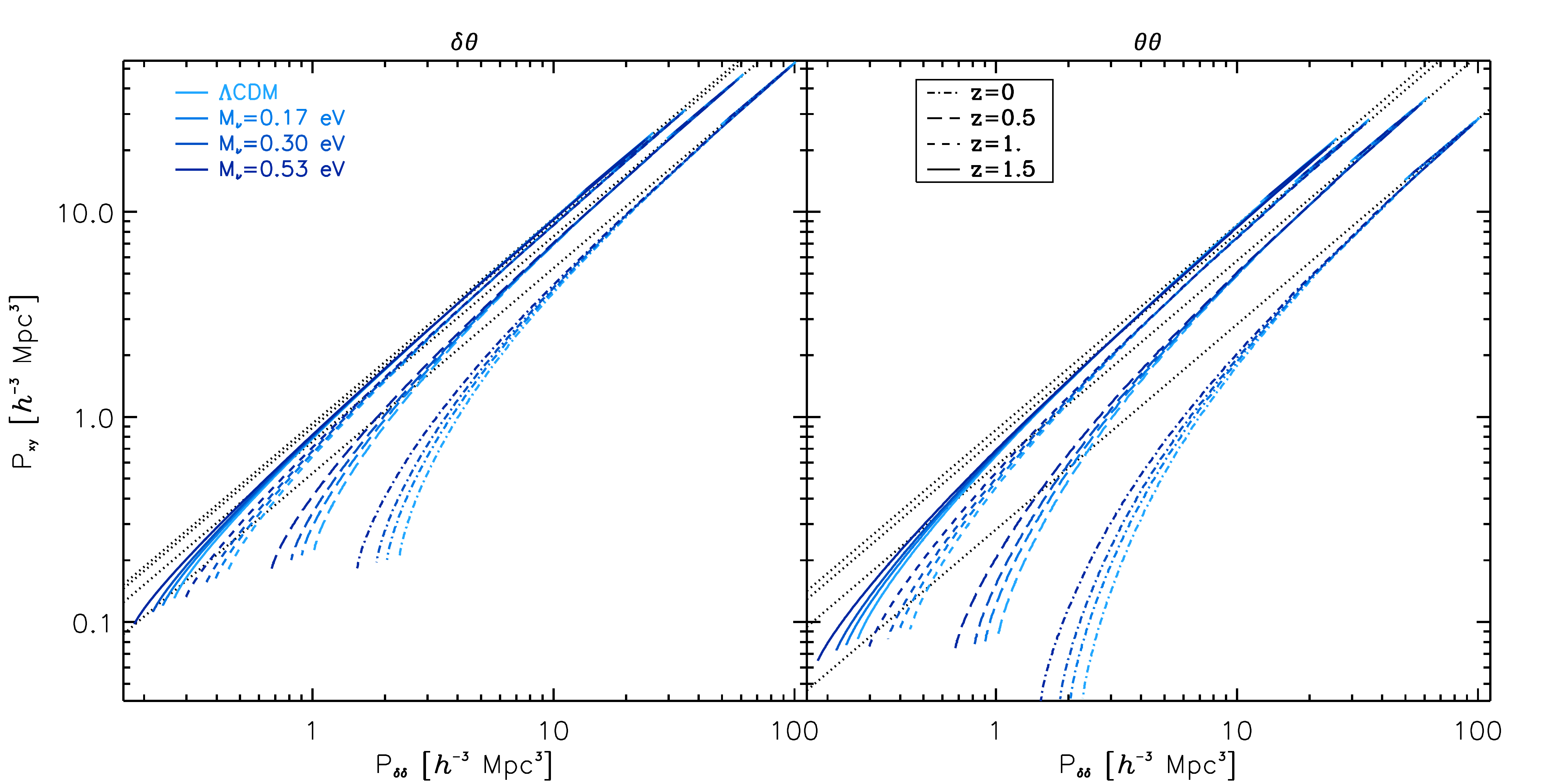}
\caption{Relation between the velocity cross (left) and auto (right) spectra with the density auto-spectrum for the different redshifts and cosmologies considered in this work. The shade of blue represents different values of the total neutrino mass, darker when the neutrino mass is increasing. The various redshifts ($z=0$, $0.5$, $1$, $1.5$) are respectively represented with dot-dashed, long dashed, short dashed, and solid lines. The black dotted line represents for each redshift the linear mapping  ($P_{\delta\theta}^{\rm Lin} = f P_{\delta\delta}^{\rm Lin}$ and $P_{\delta\theta} =  f^2 P_{\delta\delta}^{\rm Lin}$ taken in the $\Lambda$CDM limit, i.e., at small $k$). 
}
\label{fig:velocity_vs_density_spectra}
\end{figure*}

The fitting functions for the velocity power spectra adopted in  \citet{Jennings2011} and \citet{Jennings2012} describe the nonlinear velocity spectra $P_{\theta\theta}$ and $P_{\delta\theta}$ in terms of the nonlinear matter power spectrum $P_{\delta\delta}$ assuming a cosmology-independent relation between these quantities at redshift zero and introducing a scaling relation to extend the results at higher redshift. However, the relations between $P_{\theta\theta}$ and $P_{\delta\delta}$  and between $P_{\theta\delta}$ and $P_{\delta\delta}$ are not universal and depend strongly, in the first place, on the amplitude of linear fluctuations, as measured for example by $\sigma_8$. This is particularly evident when comparing our set of massive neutrino cosmologies where the neutrino mass  directly affects this quantity. In Figure \ref{fig:velocity_vs_density_spectra}, the velocity spectra (auto and cross) are plotted as a function of the corresponding matter density power spectrum. These relations, far from being universal, clearly depend as much on redshift as they depend on the sum of neutrino masses, via the amplitude suppression induced by the latter. The cosmology-independence of the fit proposed by \citet{Jennings2011} is perhaps justified by the fact that the different cosmological models considered in that paper share the same amplitude normalisation in terms of the $\sigma_8$ parameter. 

We chose a different approach to fit the velocity power spectra measured in the DEMNUni-I simulations.
In our set of simulations, all cosmological models considered present the same amplitude for the total-matter power spectrum in the large-scale limit, 
matching 
current CMB constraints \citep{Planck2016}. However, since the presence of massive neutrinos suppresses the growth of fluctuations 
below the free-streaming scale, we obtain quite different values for $\sigma_8(z=0)$ as a function of neutrino mass
(see Table \ref{table1}). We are thus able to test a wider range of amplitudes and shapes of the linear power spectrum, since neutrinos affect both of them. 

%
%
%
%
%
\begin{figure*}
\hspace{0.5cm}
\centering
\resizebox{0.95\hsize}{!}{\includegraphics{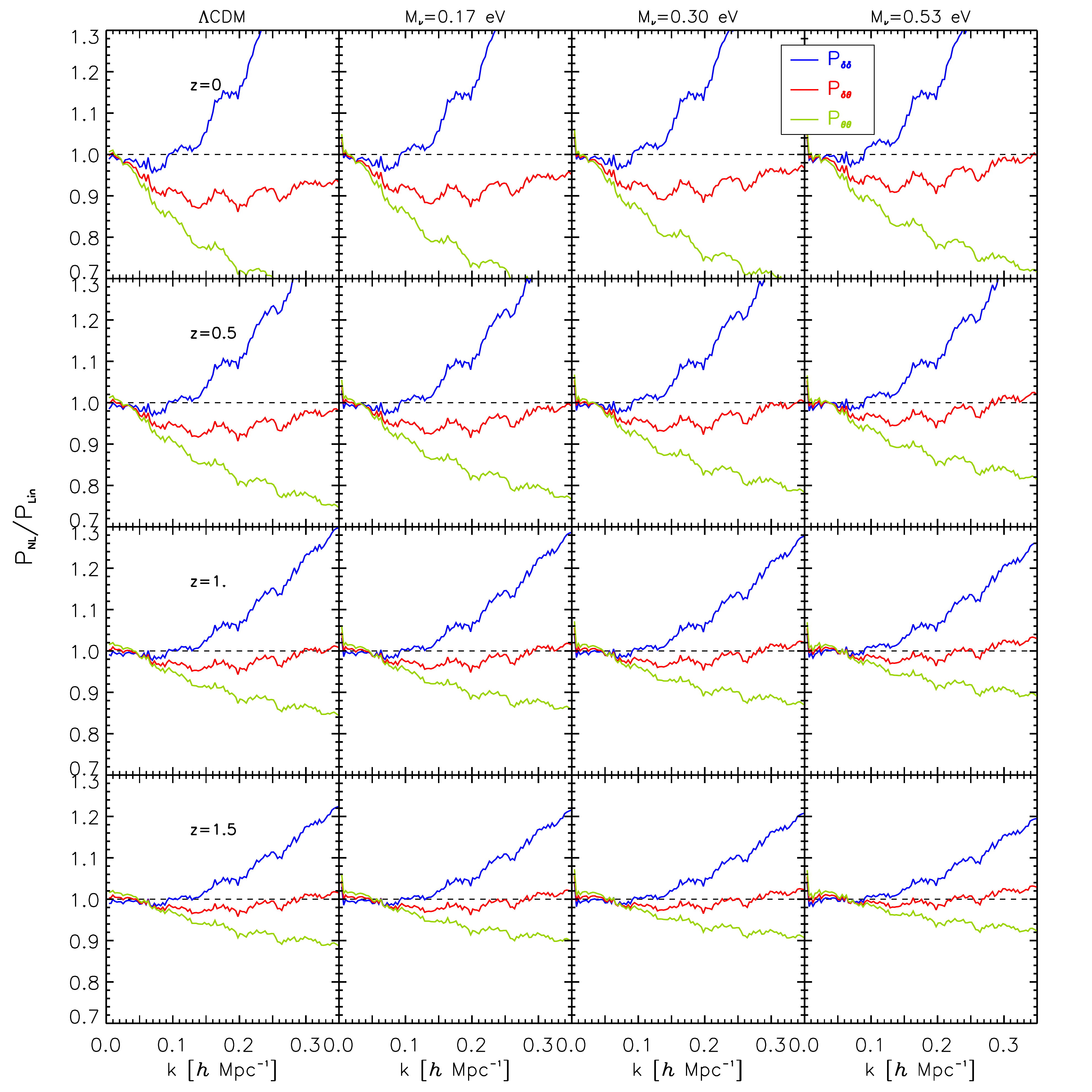}}
\caption{Ratio between nonlinear (measured) $P_{NL}$ and the linear (predicted) $P_{Lin}$ power spectra in the $\Lambda$CDM case ($\sum m_\nu=0$) and for the three neutrino masses ($\sum m_\nu=\{0.17,\; 0.3, \; 0.53\}$ eV). The density--density, density--velocity divergence, and velocity divergence--velocity divergence spectra are represented by blue solid, red short dashed, and green long dashed lines, respectively. Each row shows the redshift evolution of these ratios (from top to bottom  $z=0$, $0.5$, $1.$ and $1.5$.) 
}
\label{fig:non_linear_evolution}
\end{figure*}
%


Figure \ref{fig:non_linear_evolution} shows the ratio of the measured power spectrum of the density, velocity, and cross to their respective linear predictions. Across all cosmological models and redshifts considered, we notice the usual increase in small-scale power for the density power spectrum $P_{\delta\delta}$ as well as the slower nonlinear growth for the velocity perturbations leading to a nonlinear suppression of $P_{\delta\theta}$ and particularly of $P_{\theta\theta}$ \citep[see e.g.,][]{bernardeau02}.

Following \citet{HA&A2015}, we therefore 
 choose to model the nonlinear corrections to the velocity spectra in terms of 
damping functions, in order to account for the suppression of power 
characterising the velocity divergence field. In addition, the level of such suppression will be, as observed, dependent on cosmology. In our approach, we do not assume a universal mapping independently of the considered cosmology. The main motivation supporting this is the empirical evidence that the velocity spectra are damped differently for different cosmological backgrounds.


In this section we explain our  general fitting formulae and show how the chosen parameters depend on the value of the overall matter clustering. For simplicity, we first employ fitting functions featuring one single free parameter that accounts for the damping of the linear prediction in the nonlinear regime (see Fig. \ref{fig:non_linear_evolution}). As first approximation, one can model the velocity spectra using one single damping function such as
\begin{equation}
P_{\delta\theta}(k)=\left \{ (P_{\delta\delta}^{\rm HF}(k)P_{\theta\theta}^{{\rm Lin}}(k) \right\}^{\frac{1}{2}} e^{-\frac{k}{k_\delta}} 
\label{eq:fit_pdt}
\end{equation}
and
\begin{equation}
P_{\theta\theta}(k)=P_{\theta\theta}^{{\rm Lin}}(k)e^{-\frac{k}{k_\theta}},
\label{eq:fit_ptt_1p}
\end{equation}
where the 
(only) 
two free parameters are the typical damping scales $k_\delta$ and $k_\theta$. We note that $P_{\delta\delta}^{\rm HF}(k)$ refers to the nonlinear density-density CDM power spectrum computed from the \textit{Halofit} calibration of \citet{halofit} while $P_{\theta\theta}^{{\rm Lin}}(k)$ refers to the linear auto-spectrum of the velocity divergence which can be computed as $P_{\theta\theta}^{{\rm Lin}}(k) = f^2(k) P_{\delta\delta}^{{\rm Lin}}(k)$. 

The fit for the velocity power spectra is carried out using a least-squares approach, 
that is, we compute the likelihood of the parameters given the measured spectra with the $\chi^2$ function defined as
\begin{equation}
\chi^2=\sum_{i=1}^{N}\frac{\left [ P_{xy}^{\rm NL}(k_i)-P_{xy}(k_i) \right ]^2}{\sigma_i^2},
\end{equation}
where $N$ is the total number of wavenumbers considered in the fit, $P_{xy}^{\rm NL}(k_i)$ is the measured auto- or cross-spectrum and $\sigma_i^2$ is its variance at the $i-$th wave mode $k_i$.  We limit the fitting range to $k_{\rm max} = 0.6\; h$ Mpc$^{-1}$, and since we have only one realisation for each cosmology we neglect the shot-noise and the nonGaussian contributions to the covariance between wave modes. Under these assumptions the error can be approximated as
\begin{equation}
\sigma_i =\frac{k_F}{ \sqrt{2\pi}}\frac{P_{xy}(k_i)}{k_i}.
\end{equation}
\begin{figure*}
\centering
\includegraphics[width=\textwidth]{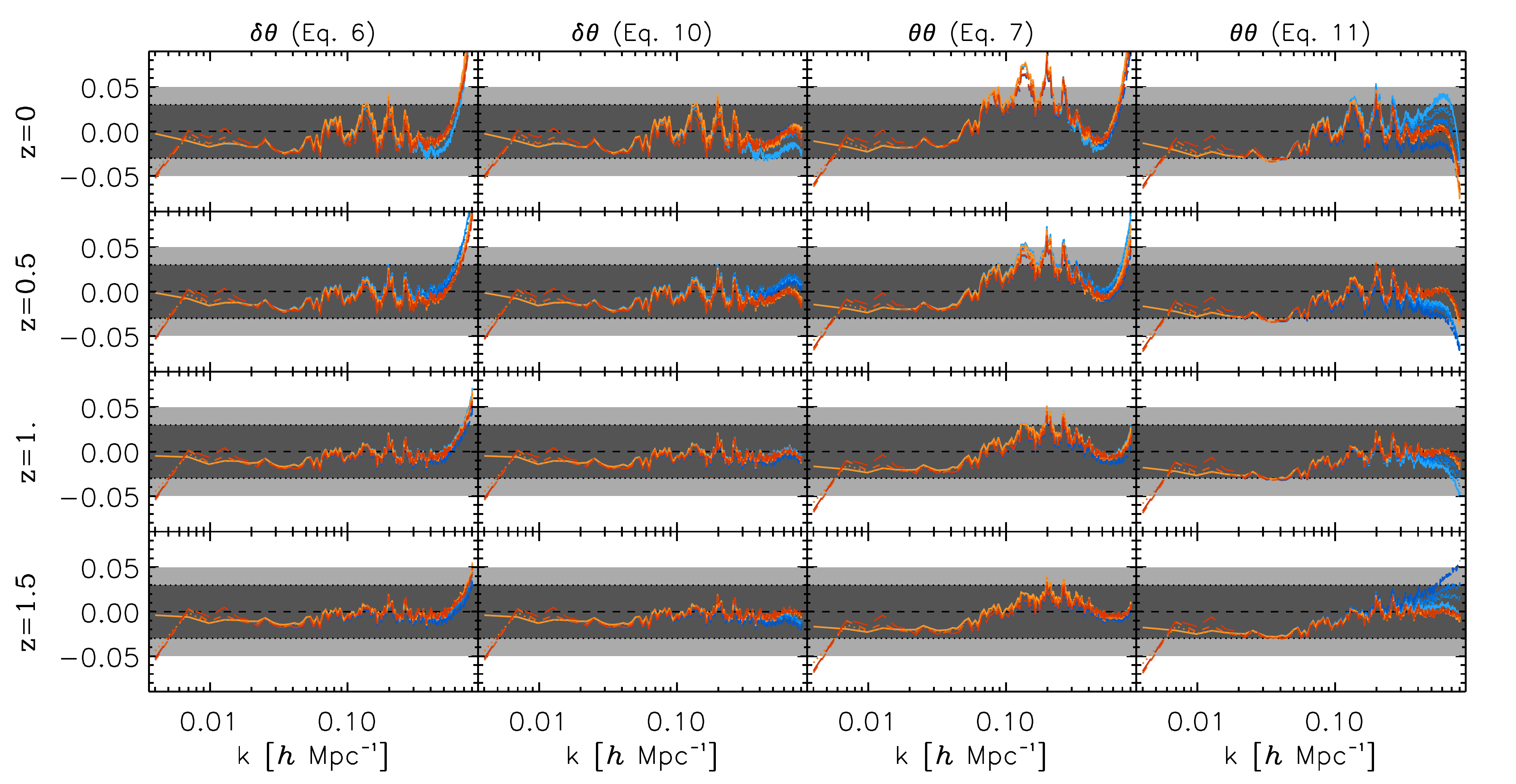}
\caption{Fractional deviation of the measured velocity spectra from our best-fitting models. Different rows mark different redshifts as specified on the left of the panels. The first and third columns show the fit of  $P_{\delta\theta}$ using equation \ref{eq:fit_pdt}, and the fit of $P_{\theta\theta}$ using equations \ref{eq:fit_ptt_1p}. The second and fourth columns show results corresponding to equations \ref{eq:fit_pdt_bis} and \ref{eq:fit_ptt_3p}. The black dashed line marks the $0$, while the light/dark gray bands represent the $5\%$/$3\%$ deviations from the measurements. Finally, the red shaded color shows the results for various neutrino masses (darker when increasing the neutrino mass) using the fitted shape parameters while the blue shaded lines show the results when using the fitted dependance of the shape parameters with respect to $\sigma_{8,m}$ (see Eqs. \ref{s8fitted}).} 
\label{fig:fit_spectra}
\end{figure*}
We note that our choice of using a fixed power $\alpha=1$ for the exponent, rather than having more degrees of freedom (i.e., $\exp(-(k/k^*)^\alpha)$), comes from a further test we carried out, which shows that the best-fit value of $\alpha$ is always close to $1$ if we treat it as a free parameter \citep[see also][]{HA&A2015}. We note that this is at odds with what one would expect when considering the propagator in renormalized perturbation theory \citep{BC&S2008,BC&S2012,CS&B2012}, in which case $\alpha$ would be closer to $2$.

The results of the fit are shown in the first and third columns of Figure \ref{fig:fit_spectra}. One can see that the simple modeling provided by Eq. \ref{eq:fit_pdt} is able to reproduce the cross power spectrum $P_{\delta\theta}$ in the nonlinear regime with an accuracy better than $5$\% up to $k=0.65\; h$ Mpc$^{-1}$. 
The fit is not working as well for the 
auto spectrum $P_{\theta\theta}$
(third panel)
especially at low redshift. The inaccuracy in this case can reach $7$-$8$\% in between  $k=0.1h/\mathrm{Mpc}$ and $k=0.3h/\mathrm{Mpc}$. Nonetheless, this approximation could be considered sufficient
for analyses that do not require precision around the BAO scale of  better than few percent. For more general applications, we improve the accuracy of the model by increasing the degrees of freedom of the fitting functions.

In the case of the cross-power spectrum, it is sufficient to add only one parameter $b$, which we fit between $k=0.55$ and $k=0.7\; h$ Mpc$^{-1}$,  
\begin{equation}
P_{\delta\theta}(k)=\left \{ (P_{\delta\delta}^{\rm HF}(k) P_{\theta\theta}^{{\rm Lin}}(k)\right\}^{\frac{1}{2}} e^{-\frac{k}{k_\delta} - bk^6} .
\label{eq:fit_pdt_bis}
\end{equation}
For the auto-spectrum $P_{\theta\theta}$, it turns out that two extra free parameters are required for a proper gain in accuracy. We therefore adopt a polynomial fitting for the damping function, as
\begin{equation}
P_{\theta\theta}(k)=P_{\theta\theta}^{{\rm Lin}}(k) e^{ -k (a_1 + a_2 k + a_3k^2) } ,
\label{eq:fit_ptt_3p}
\end{equation}
involving three parameters, which we fit in the range $0.01 <k < 0.7\; h$ Mpc$^{-1}$. The performances of the new fitting functions are shown in the second and fourth columns of Figure \ref{fig:fit_spectra}. In this case, 
the measurements of $P_{\theta\theta}$ are reproduced with a maximum systematic error of $3$\% on all scales below $k=0.7\; h$ Mpc$^{-1}$, and for all redshifts and neutrino 
masses considered.
We therefore use a total of only five free parameters $k_\delta$, $b$, $a_1$, $a_2$ and $a_3$ to mimic, with a $3$\% level accuracy at $k<0.7\; h$ Mpc$^{-1}$,  the nonlinear effects on both the auto- and cross-spectra $P_{\theta\theta}$ and $P_{\delta\theta}$.

\begin{figure}
\resizebox{\hsize}{!}{\includegraphics{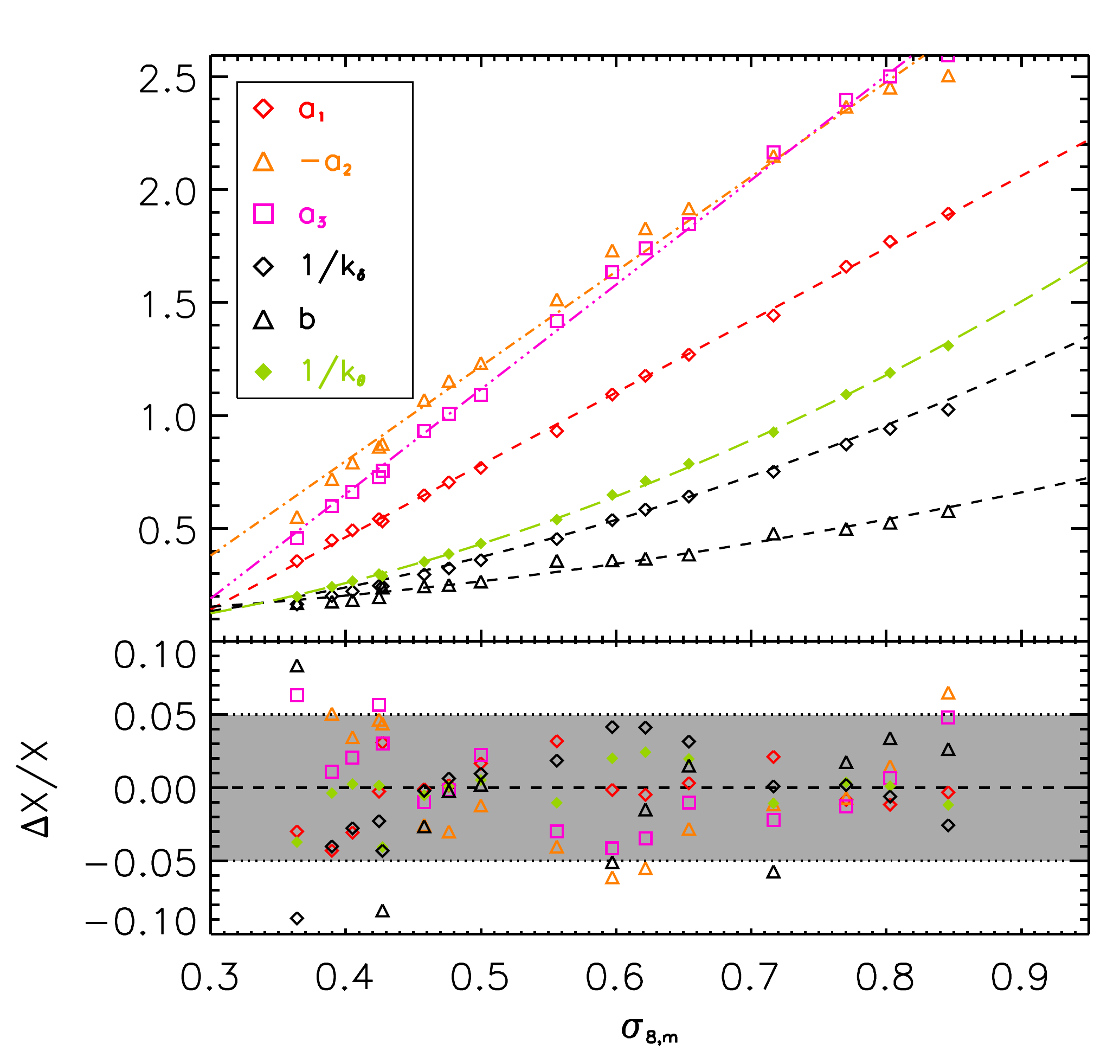}}

\caption{{\it Top:} Dependence of the fitting parameters $k_\delta$, $b$, $k_\theta$, $a_1$, $a_2$ and $a_3$ on the total matter clustering amplitude $\sigma_{8,m}$ at each redshift, in the corresponding simulation (symbols). Lines are showing the fit of the dependance. {\it Bottom:}  Residual between the fitted $\sigma_{8,m}$ dependance and the measured one.}
\label{fig:kcut_vs_sigma8}
\end{figure}

Let us now focus on the sensitivity of these parameters to cosmology and specifically to the overall amplitude of the matter power spectrum. From our set of four simulations at four different redshifts we are able to span a large range of possible values of $\sigma_8$, which we use as a proxy for the amount of nonlinearities. Regarding neutrino cosmologies it is necessary to choose whether we use the $\sigma_8$ defined for cold dark matter only, $\sigma_{8,c}$ , or the one defined for the total matter, $\sigma_{8,m}$. It has been shown \citep{Castorina2015} that regarding the bias or the nonlinear effects on the density power spectrum $P_{\delta\delta}$, what matters is the amplitude of the  cold dark matter clustering and not the total one. We have analyzed the dependency of the fitted parameters with respect to both amplitudes $\sigma_{8,c}$ and $\sigma_{8,m}$  and we found that one should use the total matter $\sigma_{8,m}$ parameter in order to assess the correct values of $k_\delta$, $b$, $a_1$, $a_2$ and $a_3$; at least, this choice is the one that lowers the residual dependency with respect to the neutrino mass. 
Figure \ref{fig:kcut_vs_sigma8} shows 
that the cosmological dependance of the fitting parameters can be mostly encapsulated into the $\sigma_{8,m}$ parameter evaluated at the corresponding redshift. The most relevant example is the good match at redshift $1.5$ of the $\Lambda$CDM simulation with the $M_\nu=0.53$eV simulation at redshift $1$; these simulations share almost the same value of the total matter $\sigma_{8,m}$, while differing significantly in the power spectrum shape.
For the two cases, we obtain the same values for the fitted parameters $k_\delta$ and $k_\theta$. 

On the contrary, when using the cold dark matter $\sigma_{8,c}$ parameter (see Table \ref{table1}) 
the residuals are
increasing because of spurious neutrino mass dependence when going from one cosmology to another. This effect can be explained by the fact  that if massive neutrinos have a weak effect on the dark matter clustering in the nonlinear regime, they add, instead, a relevant contribution to the velocity field which is felt by dark matter particles. As a result, it seems that the preferred dependence on the total matter $\sigma_{8,m}$ comes from the fact that nonlinearities in the velocity field are generated by the whole matter distribution (cold dark matter plus neutrinos). This confirms the need for running cosmological simulations including massive neutrino particles in order to generate a velocity field correctly treated in the nonlinear regime, especially for what concerns RSD analyses.
 
The final step of our fitting process is to fit the dependence of the shape parameters $k_\delta$, $b$, $a_1$, $a_2$, $a_3$ and $k_\theta$ with respect to the total matter $\sigma_{8,m}$. To this purpose we limit ourselves to linear or quadratic fitting depending on the parameters, finding
\begin{eqnarray}
a_1 & = & -0.817 + 3.198\sigma_{8,m} \,,\nonumber \\
a_2 & = & 0.877 -4.191\sigma_{8,m} \,,\nonumber  \\
a_3 & = &  -1.199  + 4.629\sigma_{8,m} \,,\nonumber \\
1/k_\delta & = & -0.017 + 1.496\sigma_{8,m}^2 \,,\nonumber \\
b & = & 0.091 + 0.702\sigma_{8,m}^2 \,,\nonumber\\
1/k_\theta & = & -0.048 +  1.917 \sigma_{8,m}^2 \,,\label{s8fitted} 
\end{eqnarray}
where $\sigma_{8,m}$ refers to the linear {\it rms} of total matter fluctuations computed at the required redshift. Therefore, the cross- and auto-spectra $P_{\delta\theta}$ and $P_{\theta\theta}$ can be computed as follows: compute the linear and nonlinear cold dark matter power spectrum at the required redshift, evaluate the linear $\sigma_{8,m}$
as 
\begin{equation}
\sigma_{8,m}^2 = \int_0^{+\infty} 4\pi k^3P_{m}(k) \hat W^2(kR) \dif \ln k,
\label{sig8}
\end{equation}
where $R=8h^{-1}$Mpc, $\hat W(x)\equiv 3/x^3\left [ \sin x - x\cos x\right]$ and $P_m$ is the linear total matter power spectrum. Finally,  compute the $k_\delta$, $b$, $k_\theta$,  $a_1$, $a_2$ and $a_3$ parameters from Eq. \ref{s8fitted} and the velocity spectra from Eqs. (\ref{eq:fit_pdt_bis}) and (\ref{eq:fit_ptt_3p}) (or \ref{eq:fit_pdt} and \ref{eq:fit_ptt_1p} depending on the required accuracy). In order to summarize the overall accuracy of our fitting formula, we show in Fig. \ref{fig:fit_spectra} the comparison between the intrinsic accuracy of the fitting formula (for the shape) in red shade and the final accuracy obtained assuming the additional fitted dependency on $\sigma_{8,m}$ in blue shade. One can see that the accuracy below $k=0.8\; h$ Mpc$^{-1}$ is about $3$\% for the cross-power spectrum while the auto-power spectrum reaches a similar accuracy below $k=0.7\; h$ Mpc$^{-1}$.

\section{Summary}
\label{sec:summary}

%
We set up an original algorithm in order to estimate the velocity field in cosmological $N$-body simulations.
From those measurements performed on sixteen particle distributions spanning four different cosmological models and four redshifts, that is, a series of $16$ snapshots, we have shown that the mapping from the nonlinear CDM density power spectrum $P_{\delta\delta}$  to the nonlinear velocity spectra $P_{\delta\theta}$ and $P_{\theta\theta}$ cannot be considered as universal  but shows a clear dependence on the amplitude of  dark matter clustering (see Fig. \ref{fig:velocity_vs_density_spectra}). 

Adopting a very simple modeling involving only two free parameters, we managed to reproduce our measurements with a precision of $3$\% below $k=0.6\; h$ Mpc$^{-1}$ for the cross-power spectrum $P_{\delta\theta}$. However, we reached only a $5$\% accuracy at redshifts higher than or equal to $z=0.5$ and for $k$ lower than $0.7\; h$ Mpc$^{-1}$. We then present an improved version of the fitting function which involves a total of five free parameters (two for the cross-spectrum and three for the auto-spectrum) and allows us to obtain an overall accuracy of $3$\% for wave modes lower than $0.7\; h$ Mpc$^{-1}$. 

Finally, we we showed evidence of the dependence of the shape parameters of the proposed fitting functions 
on the total matter \textit{r.m.s.} fluctuations in spheres of radius $8h^{-1}$Mpc and proposed simple fitting forms for the shape parameters with respect to the $\sigma_{8,m}$ parameter evaluated at the considered redshift. 
This preferred dependence with respect to the total matter amplitude rather than the cold dark matter one confirms the relevance of the neutrino perturbations on the cold dark matter velocity field in the nonlinear regime.

In a future paper we shall generalize these results using the second set of the DEMNUni cosmological simulations that include dynamical 
dark energy,
extending 
our study of the dependence of the shape parameters 
on the {\it r.m.s.} amplitude of clustering 
$\sigma_{8,m}$. 


\section*{Acknowledgements}

This work were developed in the framework of the ERC \textit{Darklight} project, supported by an Advanced Research Grant of the European Research Council (n. 291521). CC, LG and ES also acknowledge a contribution from ASI through grant I/023/12/0. ES is also supported by INFN INDARK PD51 grant. The DEMNUni-I simulations were carried out in the framework of the ``The Dark Energy and Massive-Neutrino Universe" using the Tier-0 IBM BG/Q \textit{Fermi} machine  of the \textit{Centro Interuniversitario del Nord-Est per il Calcolo Elettronico} (CINECA). We acknowledge a generous CPU allocation by the 
Italian SuperComputing Resource Allocation (ISCRA).
\bibliographystyle{aa}
\bibliography{velocity.bib}

\end{document}